%% file: emw.tex
\documentclass[aps,prb,amsmath,amssymb,twocolumn,floatfix,10pt,footinbib,showpacs]{revtex4-1}
\usepackage{graphicx}
\usepackage{graphics,color}
\usepackage{hyperref}

\newcommand{\bs}{\;\;\;\;\;}

\begin{document}

\title{Exact diagonalization study of the tunable edge magnetism in graphene}
\author{David J. Luitz}
\author{Fakher F. Assaad}
\affiliation{Institute for Theoretical Physics and Astrophysics, University of W\"urzburg, 
Am Hubland, 97074 W\"urzburg, Germany}
\author{Manuel J. Schmidt}
\affiliation{Department of Physics, University of Basel, Klingelbergstrasse 82, 4056 Basel, 
Switzerland}
\date{\today}


\begin{abstract}
The tunable magnetism at graphene edges with lengths of up to 48 unit cells is analyzed 
by an exact diagonalization technique. For this we use a generalized interacting 
one-dimensional model which can be tuned continuously from a limit describing graphene 
zigzag edge states with a ferromagnetic phase, to a limit equivalent to a Hubbard chain, 
which does not allow ferromagnetism. This analysis sheds light onto the question why the 
edge states have a ferromagnetic ground state, while a usual one-dimensional metal does 
not. Essentially we find that there are two important features of edge states: (a) umklapp 
processes are completely forbidden for edge states; this allows a spin-polarized ground 
state. (b) the strong momentum dependence of the effective interaction vertex for edge 
states gives rise to a regime of partial spin-polarization and a second order phase 
transition between a standard paramagnetic Luttinger liquid and ferromagnetic Luttinger 
liquid.
\end{abstract}

\maketitle

\section{Introduction}
\input{introduction}

\section{Edge state models\label{model_section}}
\input{models}

\section{Exact diagonalization\label{ed_section}}
\input{exact_diag}

\section{Discussion\label{discussion}}
\input{discussion}

\acknowledgments

D.J.L. and F.F.A. acknowledge financial support from the DFG for grant AS120/4-3. M.J.S. acknowledges financial support from the Swiss NSF and from the NCCR QSIT. 

\appendix

\section{Variational analysis of the generalized model\label{appendix_mean_field}}
\input{mean_field}

\section{Exact diagonalization of the direct model\label{appendix_ed_original_model}}
\input{direct_model_ed}


\bibliography{../paper_bosonization/tem_bib.bib,emw.bib}

\end{document}

%% file: introduction.tex
Since it has first been isolated in the laboratory,\cite
{novoselov_graphene_first_realization} graphene, a two-dimensional honeycomb lattice 
of carbon atoms,\cite{graphene_review_1} attracts much attention. In fact, graphene has 
multiple amazing properties. To name only a few of them, it ranges among the mechanically 
strongest materials,\cite{lee_graphene_mechanical_strength} it shows a quantum 
Hall effect at room temperature,\cite{zhang_graphene_qhe} and, due to its unusual 
Dirac band structure, it allows the study of relativistic quantum physics in a solid state 
environment.\cite{novoselov_dirac_femions_graphene} Furthermore, its potential 
application as the basis of the next generation of electronic devices stimulated great 
efforts to gain experimental control as well as theoretical understanding of this 
astonishing material.

Usually a strong electron confinement increases the strength of electron-electron 
interactions by pushing the electrons close together. However, in spite of the extreme 
electron confinement to only one single layer of atoms, many experiments in graphene 
may be explained by assuming the electrons to be non-interacting. 
This is especially true for experiments probing the bulk properties of graphene, as the bulk density
of states vanishes at the Fermi level, suppressing the manifestation of interaction effects.
On the other hand, the properties of zigzag edges differ greatly from the bulk properties. So-called edge states, i.e. one-dimensional states with very small bandwidth, localized at these edges, give rise to a peak in the local density of states at the Fermi
energy.\cite{edge_states_fujita_1996}  The enhanced density of states allows the electron-electron
interaction to drive the zigzag edges to a ferromagnetic state with a magnetic moment localized at
the edge.\cite {son_abinitio_prl_2006, son_half_metallic_gnrs_2007, jung_em_mean_field_2009,
feldner_qmc_2010,feldner_11, bosonization_dmrg_hikihara_2003} This phenomenon is known as edge magnetism.

At normal graphene edges the electron-electron interaction is so strong and the 
bandwidth of the edge states is so small that the spins of all electrons in the edge states 
are completely aligned. However, as has been proposed recently, graphene/graphane 
interfaces provide means to tune the bandwidth of the edge states to regimes in which 
the edge starts to depolarize and the edge magnetism is gradually suppressed until, for 
a critical edge state bandwidth, the magnetism disappears.\cite
{tem_schmidt_loss_2010}

In Ref. \onlinecite{tem_schmidt_loss_2010} it was argued that this interaction-induced magnetism
can be understood on the basis of an effective model, describing the interacting one-dimensional edge states only, while the bulk states are neglected.
What at first glance appears to be a contradiction to the Lieb-Mattis theorem,\cite{lieb_mattis}
stating that the ground state of interacting electrons in one dimension cannot be spin polarized,
can be resolved by noting that the effective edge state model does not fulfill the prerequisites
of the Lieb-Mattis theorem.\cite{tem_schmidt_loss_2010} The deeper reasons for the existence of a ferromagnetic ground state in a one-dimensional interacting electron system, however, remained elusive. The present work is devoted to this issue.

In this paper we present a systematic exact diagonalization analysis of
interacting edge states. Two 
striking features of edge states turn out to be most important for their magnetic properties: 
(a) the edge states exist only in a restricted part of the Brillouin zone and (b) the 
transverse edge state wave function has a strong characteristic momentum 
dependence. These features have consequences for the effective low-energy theory, namely (a) \emph{no umklapp processes are allowed in the interaction Hamiltonian} and (b) \emph{the 
interaction vertex acquires an unusually strong momentum dependence}. In order to be 
able to study the consequences of these two features, we introduce a 
generalized model in which we add an artificial interaction term describing umklapp processes and allow the 
momentum-dependence of the interaction vertex to be tuned from a momentum-independent vertex, as in
usual metals, to the full momentum-dependence, as it is found 
in edge states. Therefore, the generalized model can be tuned continuously from the 
limit in which it describes edge states to a limit which corresponds to usual 
one-dimensional metals such as the Hubbard chain. We solve this generalized model for 
graphene zigzag edges of finite length $L=48$ unit cells
(i.e. $\sim$12 nm) by exact 
diagonalization using the Lanczos method for the determination of the ground state of the effective
model.\cite{lanczos_50,dagotto_rev_94,senechal_lecture_notes_08,large_system}

The paper is organized as follows. In Sec. \ref{model_section} we review the direct 
model as it has been derived in Ref. \onlinecite{tem_schmidt_loss_2010} and introduce 
the more versatile generalized model with additional tunability. In Sec. \ref{ed_section} 
the exact diagonalization analysis of the generalized model is presented. Finally, the 
results are discussed in Sec. \ref{discussion}.

%% file: models.tex
In this section we introduce the models on which our analysis is based. The edge state model
obtained from the direct projection of the honeycomb lattice Hubbard model to the Fock space
spanned only by the edge states has been discussed in Ref. \onlinecite{tem_schmidt_loss_2010}. This
model will be called the {\it direct model} in the following. We identify two important features of
the direct model: (a) the restriction of the Brillouin zone for the edge states and (b) the strong
dependence of the transverse localization length on the momentum along the edge. After having
analyzed the consequences of these features for the effective interaction vertex, we propose a
generalized edge state model in which these features can be tuned. This allows us to investigate the
impact of each of these edge state features on the magnetic properties. In particular, the
generalized model can be tuned continuously from a Hubbard chain limit, i.e., a usual one-dimensional
metal without any ferromagnetic ground state, to the edge state limit with its ferromagnetic ground
state.

\subsection{Direct derivation from the honeycomb model}
\begin{figure}[!ht]
\centering
\includegraphics[width=\columnwidth]{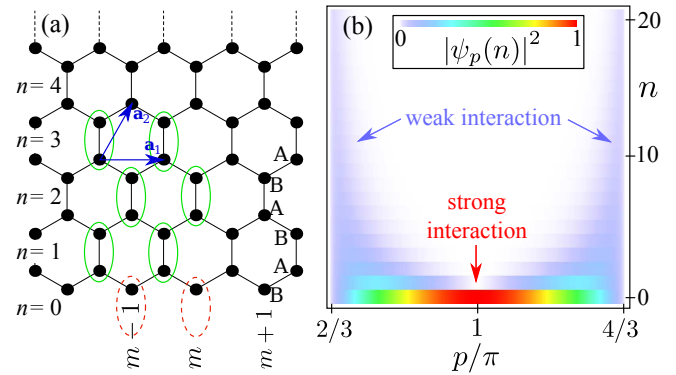}
\caption{(Color online) Part (a) shows the half-infinite honeycomb lattice. The solid elipses
(green) are the complete unit cells in the bulk region and the dashed elipses indicate the cut unit
cells at the $\alpha$ edge. The $n$ and $m$ directions are indicated as well as the sublattice
indices A, B. Part (b) shows the modulus square of the transverse (in $n$ direction) edge state wave
function $|\psi_p(n)|^2$ on the B sublattice sites in the reduced Brillouin zone $\frac{2\pi}3\leq p
\leq \frac{4\pi}3$. The extreme momentum dependence of the localization of $\psi_p$ in $n$ direction is crucial for
the magnetic properties of edge states.}
\label{fig_definitions}
\end{figure}

We start from the simplest possible non-interacting tight-binding model of electrons in graphene
zigzag ribbons, taking into account only nearest neighbor hoppings of $\pi$ electrons $\mathcal H =
\sum_{\left<i,j\right>,\sigma} c^\dagger_{i \sigma} c_{j\sigma}$, where $\left<i,j\right>$ runs over
nearest neighbor sites of a half-infinite honeycomb lattice, $i\equiv(m,n,s)$ is a collective site
index for the $(m,n)$th unit cell and the $s=A,B$ sublattice (see Fig. \ref{fig_definitions}), and
$c_{i\sigma}$ annihilates an electron at site $i$ with spin $\sigma$. Since we are exclusively
interested in the zero energy eigenstates of $\mathcal H$, the actual energy scale of $\mathcal H$
is unimportant so that we may drop it.\footnote{It is only important to assume that this energy scale
is large enough so that the restriction to the zero energy sector of $\mathcal H$ is well justified.
The validity of this assumption has been checked in Ref. \onlinecite{tem_schmidt_loss_2010}.} The
zero energy states are created by the fermionic edge state operator
\begin{equation}
e^\dagger_{p\sigma} =\sum_n \psi_p(n) c^\dagger_{pn\sigma}, \bs \psi_p(n) = \mathcal N_p u_p^{n}
\label{eq_edge_state_operator}
\end{equation}
where $u_p = -1-e^{ip}$, $p$ is the momentum in $m$ direction (along the edge), and
$c_{pn\sigma}=L^{-\frac12} \sum_{m} e^{-i p m} c_{(m,n,B)\sigma}$, with the number of unit cells $L$
in $m$ direction, i.e., along the edge. The $p$-dependent normalization constant $\mathcal N_p
= \sqrt{2\cos(p-\pi)-1}$ can be interpreted as the weight of the edge state wave function right at the edge atoms
where $n=0$.  It is easily seen that $\mathcal H e^\dagger_{p\sigma} = 0$.
As the edge state wave function is only non-zero on the $B$ sublattice we omit the sublattice index,
setting it to $s=B$.

The two most important features of the edge state wave function [Eq. (\ref{eq_edge_state_operator})]
are : (a) The edge state only exists for momenta $\frac{2\pi}3 < p < \frac{4\pi}3$. In the rest
of the Brillouin zone the edge state wave function is not normalizable, as $|u_p|>1$ for these
momenta. (b) In $n$ direction the edge state is sharply localized at the edge for $p=\pi$, wereas
for $p$ close to one of the Dirac points $K=\frac{2\pi}3$ and $K'=\frac{4\pi}3$, the wave function
delocalizes into the bulk [see also Fig. \ref{fig_definitions}(b)]. These two edge state properties
are stable against adding more details, such as second-nearest neighbor hopping or various edge
passivations, to the honeycomb Hamiltonian $\mathcal
H$.\cite{tem_schmidt_loss_2010,soi_schmidt_loss_2010} The detailed analysis presented in this paper
will clarify that \emph{the existence of edge magnetism and in
particular its tunability are consequences of these two edge state properties.} 

The $p$-dependence of the localization length of $\psi_p$ has consequences for the edge states'
self-energy $\epsilon_0$ as well as for their interaction vertex function $\Gamma$. Neglecting the bulk
state contributions,\cite{tem_schmidt_loss_2010} the self-energy correction due to a perturbation
$V$, which is invariant along the edge, is given by $\epsilon_0(p) = \left<\psi_p|V|\psi_p\right>$.
Due to the delocalization of $\psi_p(n)$ for $p$ near $K,K'$, edge-localized perturbations $V$ lead
to self-energy corrections for which $\epsilon_0(K)=\epsilon_0(K')=0$ while $\epsilon_0(\pi)\sim
||V||$. For sufficiently well behaved perturbations the self-energy correction gives rise to a
smooth edge state energy dispersion with a bandwidth $\Delta \sim ||V||$. For a large class of these
edge-localized perturbations, the self-energy correction approximately has the form
\begin{equation}
\epsilon_0(p) \simeq \mathcal N_p^2 \Delta,\label{self_energy_correction} 
\end{equation}
with $\mathcal N_p^2$ the $p$-dependent weight of the edge state wave function right at the edge.
Eq. (\ref{self_energy_correction}) expresses that an edge state which is more localized at the edge
experiences a stronger self-energy correction from an edge-localized perturbation than an edge state
which is delocalized into the bulk region. Examples of such perturbations are edge passivations,
graphane termination,\cite{soi_schmidt_loss_2010} or local interactions with the substrate. Note
that the edge state bandwidth $\Delta$ is experimentally tunable in various
ways so that we consider $\Delta$ as a free parameter. Therefore, the noninteracting part of the
direct model (dm) edge state Hamiltonian is given by:
\begin{equation}
  H_0^\text{dm} = -\Delta \sum_\sigma \sideset{}{'}\sum_p \mathcal{N}_p^2 e_{p,\sigma}^\dagger
  e_{p,\sigma},
  \label{eq:direct_model_H0}
\end{equation}
where the sum is restricted such that only edge state operators $e_{p\sigma}$ with $\frac{2\pi}3
\leq p \leq \frac{4\pi}3$ appear. 

The effective interaction of the edge states, derived by projecting the Hubbard Hamiltonian on the
two-dimensional honeycomb lattice $H_U = U \sum_i c^\dagger_{i\uparrow} c_{i\uparrow}
c_{i\downarrow}^\dagger c_{i\downarrow}$ to the Fock space spanned by the edge states,
reads\cite{tem_schmidt_loss_2010}
\begin{equation}
  H_1^\text{dm} = \frac UL \sideset{}{'}\sum_{p,p',q} \Gamma(p,p',q) e^\dagger_{p+q\uparrow}
e_{p\uparrow}e^\dagger_{p'-q\downarrow} e_{p'\downarrow}. \label{eq_effective_interaction_trig}
\end{equation}
Together, we have the effective Hamiltonian $H^\text{dm} =
H_0^\text{dm} + H_1^\text{dm}$.
The interaction vertex is given by the overlap of the wave
functions of all four fermions (with momenta $p+q,p,p'-q,p'$) participating in the interaction
\begin{multline}
\Gamma(p,p',q) = \sum_{n=0}^\infty \psi^*_{p+q}(n)\psi_p(n) \psi_{p'-q}^*(n) \psi_{p'}(n) \\=
\frac{\mathcal N_{p+q} \mathcal N_p \mathcal N_{p'-q} \mathcal N_{p'}}{1-u^*_{p+q} u_p u^*_{p'-q}
u_{p'}}.\label{eq_vertex_function}
\end{multline}
While the denominator, resulting from the geometric series over $n$, turns out to lead only to
unimportant quantitative corrections, the numerator of $\Gamma$, which is the product of the wave
function weights at the edge $\mathcal N_p$ for each of the four fermion operators, leads to the
momentum-dependence of the interaction strength which is important for the stability of the weak
edge magnetism. Essentially, the effective interaction becomes stronger the more localized the
participating fermions are, i.e., the closer their momenta are to $p=\pi$. If one or more of the
momenta are close to the Dirac points $\frac{2\pi}3,\frac{4\pi}3$, where the edge state wave
functions delocalize into the bulk, the effective interaction is suppressed (see Fig.
\ref{fig_definitions}). Note that setting the denominator in Eq. (\ref{eq_vertex_function}) to unity
corresponds to assuming that the Hubbard interaction is only present at the outermost line of carbon
atoms right at the edge. Such an approximation has been used in Ref.
\onlinecite{hohenadler_qshi_2011}. We find that this approximation is inessential for the edge
magnetism, leading only to quantitative corrections.

An important consequence of the restriction of the $p$ summation in Eq.
(\ref{eq_effective_interaction_trig}) is the {\it absence of umklapp processes}. As explained above,
edge states only exist in one third of the Brillouin zone, i.e. for $\frac{2\pi}3\leq p\leq
\frac{4\pi}3$, so that no four fermion process with momentum $\pm 2\pi$ exists. Indeed, within the
restricted Brillouin zone, the process with the largest possible total momentum $p_{\rm tot}$ is
$e^\dagger_{4\pi/3\uparrow} e_{2\pi/3\uparrow} e^\dagger_{4\pi/3\downarrow} e_{2\pi/3\downarrow}$,
i.e. $p_{\rm tot}=\frac{4\pi}3<2\pi$. Processes with larger total momentum leave the restricted
Brillouin zone and are therefore suppressed, as they involve the overlap of edge states and bulk states, which is
small. Also, most of the bulk states live in a different energy regime than the edge
states.\cite{tem_schmidt_loss_2010}

Thus, we have identified two properties of edge states which make them fundamentally different from
usual one-dimensional conductors:
\begin{enumerate}
\item[(a)] Due to the restricted Brillouin zone, umklapp processes are forbidden.
\item[(b)] The transverse localization $\mathcal N_p$ of the edge state wave function gives rise to
a tunable band width $\epsilon_0(p) \simeq \mathcal N_p^2 \Delta$. Furthermore, the interaction
vertex becomes weaker if the momenta of the participating fermions approach a Dirac point, i.e.
$\Gamma(p,p',q) \propto \mathcal N_{p+q} \mathcal N_p \mathcal N_{p'-q} \mathcal N_{p'}$.
\end{enumerate}
We will show that these properties are the basis for the magnetism at graphene edges.

\subsection{Generalized model}
\begin{figure}[!ht]
\centering
\includegraphics[width=\columnwidth]{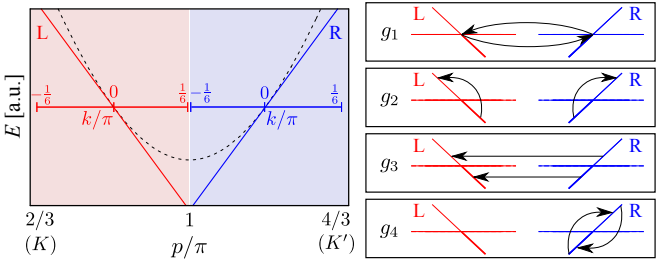}
\caption{(Color online) Left: The self-energy $\sim \mathcal N_p^2$ of the direct model (dashed
line) and the linearized self-energy (solid lines). Right: The four possible interaction processes.
The $g_3$ process is not allowed for graphene edge states.}
\label{fig_linearization}
\end{figure}

We now introduce a generalized model in which the different aspects of the effective
electron-electron interaction, found in the previous subsection, may be tuned independently. For
this, we map the edge state operators $e_{p,\sigma}$ which correspond to right (left) moving modes
for $\pi<p<\frac{4\pi}3$ ($\frac{2\pi}3<p<\pi$), to fermionic operators $c_{kr\sigma}$ in which
$r=R,L$ specifies the direction of motion and $-\frac\pi6 \leq k \leq \frac\pi6$, i.e. $c_{kr\sigma}
= e_{k+\pi+r\pi/6,\sigma}$ and $p=k+\pi+r\pi/6$. The direction of motion $r=R,L$ corresponds to
$r=\pm1$ when used in formulas. Note that the zero point of $k$ has been shifted so that $k=0$
corresponds to $p=\pi\pm\pi/6$ for right and left movers, respectively (see Fig.
\ref{fig_linearization}).

For the non-interacting part of the generalized edge state model we assume a linear spectrum with
slope $\pm v_F$
\begin{equation}
H_0 = v_F \sum_{\substack{r=R,L\\\sigma=\uparrow,\downarrow}} \sum_{k=-\pi/6}^{\pi/6} (r k)
c_{kr\sigma}^\dagger c_{kr\sigma}.
\end{equation}
This linearization of the self-energy [Eq. (\ref{self_energy_correction})] only leads to inessential
quantitative corrections (see also Appendix \ref{appendix_ed_original_model}).

The partitioning into left- and right-movers always gives rise to four terms in the interaction part
$H_1$ of the Hamiltonian involving different combinations of left- and right-moving modes.
Conventionally, these terms are called $g_1,g_2,g_3,g_4$ processes (see Fig. \ref{fig_linearization}
and Ref. \onlinecite{giamarchi_book}). $g_2,g_4$ correspond to forward scattering, involving
processes that scatter only between modes with the same direction of motion, $g_1$ refers to
backward scattering, and $g_3$ are the umklapp terms which are forbidden in edge states. Note that,
unlike in usual $g$-ology,\cite{giamarchi_book} we may not assume that the coupling constants for the individual $g_i$
processes are constant. The momentum-dependence of the $g_i$ must be taken into account.

The two forward scattering processes $g_2$ and $g_4$ may be merged together into one
Hamiltonian\footnote{The equal strength of the $g_2$ and $g_4$ processes is due to the symmetry
property $\Gamma(-k_F,k_F,2k_F) = \Gamma(\pm k_F,\pm k_F,0)$, which is a consequence of the origin
of $\Gamma$ in the 2D honeycomb Hubbard model (see Ref. \onlinecite{tem_schmidt_loss_2010}).}
\begin{multline}
H^{\rm fs}_1 = \frac UL \sum_{r,r'} \sideset{}{'}\sum_{k,k',q} S^r_{k+q} S^r_k S^{r'}_{k'-q}
S^{r'}_{k'}\times \\
:c^\dagger_{k+qr\uparrow} c_{kr\uparrow} c^\dagger_{k'-q r'\downarrow}
c_{k'r'\downarrow}:,\label{def_h_fs}
\end{multline}
where $:A :$ enforces the normal order\cite{giamarchi_book} of the operator $A$. The primed sum is
restricted such that $|k|<\pi/6$ for all momentum arguments $k$ in the electron operators. In order
to be able to change the amplitude of the momentum-dependence of the interaction vertex
$\Gamma(p,p',q)$ [see Eq. (\ref{eq_vertex_function})], we introduce the factors
\begin{equation}
S_k^r = \sqrt{1-r\Gamma_1 k},
\end{equation}
from which we build the interaction vertex for the generalized model. The factor
$\Gamma_1\in[0,6/\pi]$ quantifies the momentum-dependence. For $\Gamma_1=0$ the interaction is
momentum-independent. This limit corresponds to usual one-dimensional Hubbard chains. For
$\Gamma_1=6/\pi$ the interaction goes to zero if at least one of the fermions is close to the upper
band edge ($k=r\pi/6$). This corresponds to the direct model, where the trigonometric term under the
square root in $\mathcal N_p$ has been replaced by a linear approximation. The differences between
the generalized model in the edge state limit and the direct model only lead to quantitative
renormalizations of the critical point, as shown in Appendix \ref{appendix_ed_original_model}. The
essential property of the interaction vertex is that it approaches zero if one of the fermion
momenta gets close to the Dirac points. This feature is present in the direct and in the generalized
model with $\Gamma_1=6/\pi$.

The form of the backscattering $(g_1)$ Hamiltonian $H_1^{\rm bs}$ is similar to $H_1^{\rm fs}$.
However the scattering takes place between left- and right-movers
\begin{multline}
H^{\rm bs}_1 = \lambda_{\rm bs} \frac{U}L \sum_r \sum_{k,k',q} S^r_{k+q} S^{-r}_k S^{-r}_{k'-q}
S^r_{k'} \times\\c^\dagger_{k+q,r,\uparrow} c_{k,-r,\uparrow} c^\dagger_{k'-q,-r,\downarrow}
c_{k',r,\downarrow}.\label{eq_ham_backscattering}
\end{multline}
We have introduced the additional parameter $\lambda_{\rm bs}$ which allows us to tune the overall
strength of the $g_1$ processes relative to the $g_2,g_4$ processes. $\lambda_{\rm bs}=1$
corresponds to the physical backscattering strength which is required by SU(2)
invariance.\cite{giamarchi_book} Nevertheless, we will investigate the consequences of a suppression
of backscattering since this will be important for a bosonization analysis of the generalized model
which will be presented in an upcoming paper.\cite{schmidt_bosonization_em_11}

As already pointed out, an important feature of edge states is the absence of umklapp processes in the effective electron-electron interaction.
However, in order to be able to compare the edge state model to a Hubbard chain, we add an
artificial umklapp process with relative strength $\lambda_{\rm us}$ to the Hamiltonian of the
generalized model \begin{multline}
H^{\rm us}_1 = \lambda_{\rm us} \frac UL \sum_r \sum_{k,k',q} S^r_{k+q} S^{-r}_k S^{r}_{k'-q}
S^{-r}_{k'} \times\\c^\dagger_{k+q,r,\uparrow} c_{k,-r,\uparrow} c^\dagger_{k'-q,r,\downarrow}
c_{k',-r,\downarrow}.
\end{multline}
Varying $\lambda_{\rm us}$ between 1 (Hubbard chain limit) and 0 (edge state limit) allows us to
investigate the consequences of the presence of umklapp processes for one-dimensional ferromagnetism.

Alltogether, the four parameters $v_F/U$, $\lambda_{\rm bs}$, $\lambda_{\rm us}$, and $\Gamma_1$
define the phase space of the generalized model
\begin{equation}
H = H_0 + H_1^{\rm fs} + H_1^{\rm bs} + H_1^{\rm us}.\label{total_hamiltonian}
\end{equation}
The following limits of this model may be identified:
\begin{enumerate}
\item Edge state limit: the generalized model with the parameters $\lambda_{\rm bs}=1$,
$\lambda_{\rm us}=0$, and $\Gamma_1 = 6/\pi$, is a good approximation of the direct model.
\item Hubbard chain limit: for $\lambda_{\rm bs}=1$, $\lambda_{\rm us}=1$, and $\Gamma_1 =0$, the
generalized model essentially describes a one-dimensional Hubbard chain. The only difference is the
assumption of a linearized single-particle spectrum instead of the $2\cos(k)$ dispersion.
\end{enumerate}

Note that it is important to work in the k-space formulation because it is difficult to control the umklapp scattering or the momentum dependence of the interaction vertex in a real space formulation. One reason for this is that an interaction vertex $\Gamma(p,p',q)$ with a nontrivial $p,p'$ dependence does not transform to a real space interaction of the form $V(x-x')$ but to a complicated non-local interaction. This also hampers the application of DMRG methods to this problem.

%% file: exact_diag.tex
The ground state of the generalized model is calculated for finite sized zigzag edges up to $L=48$ by 
the Lanczos exact diagonalization method\footnote{An edge with $L$ unit cells in length corresponds to only 
$L/3$ k-space points in the reduced Brillouin zone in which the edge states are
defined.}\cite{large_system,dagotto_rev_94,lanczos_50,senechal_lecture_notes_08}. The magnetic 
properties of the ground state depend on the ratio between the kinetic energy and the potential energy 
$v_F\pi/ U$, which is experimentally tunable at graphene/graphane interfaces.\cite
{tem_schmidt_loss_2010} Three additional tuning parameters $\Gamma_1,\lambda_{\rm us},\lambda_
{\rm bs}$, which are not accessible experimentally, have been added in order to be able to study the 
significance of the momentum dependence of the interaction vertex ($\Gamma_1$), the influence of 
the absence of umklapp scattering ($\lambda_{\rm us}$) in edge states, and also the importance of 
backscattering ($\lambda_{\rm bs}$). With those artificial parameters, the generalized model may be 
tuned continuously from a Hubbard chain limit to the edge state limit. In both limits the model describes 
an interacting one-dimensional metal. The magnetic properties in these two limits, however, differ 
strongly: while the usual Hubbard chain (with umklapp scattering and without momentum dependence) 
does not give rise to a ferromagnetic ground state, the edge states (without umklapp scattering and 
with momentum-dependent interactions) show two magnetic phases in addition to the non-magnetic 
Luttinger liquid phase: for strong interactions the saturated edge magnetism\cite
{jung_em_mean_field_2009, son_abinitio_prl_2006, son_half_metallic_gnrs_2007} is recovered, while 
for intermediate interaction strengths, a ferromagnetic Luttinger liquid appears.

The Hamiltonian $H$ [Eq. (\ref{total_hamiltonian})] conserves the numbers $N_\uparrow,N_\downarrow
$ of up-spin and down-spin electrons, so that $H$ is block diagonal in the $S_z$ subspaces, which we 
define by the total spin-polarization in $z$ direction
\begin{equation}
S_z = \frac12(N_\uparrow - N_\downarrow) = 0,1,2,...,N/2.
\end{equation}
The total number of electrons $N=N_\uparrow+N_\downarrow = L/3$ is kept constant. This 
corresponds to half-filling. Note, however, that the filling is physically relevant only if umklapp scattering 
is present (i.e. $\lambda_{\rm us}>0$). For the edge states in which we are finally interested, umklapp 
scattering is forbidden so that the filling is irrelevant as it only leads to quantitative renormalizations of 
the interaction strength and the Fermi velocity. In the following, we determine the ground state of
$H$ in each $S_z$ subspace separately.

Note that by the definition of the $S_z$ subspaces we have chosen a spin quantization axis. The 
Hamiltonian $H$, however, is SU(2) symmetric if the backscattering is at its physical strength $
\lambda_{\rm bs}=1$. Furthermore, since we are dealing with finite systems, there will be no 
spontaneous rotational symmetry breaking. Thus, edge magnetism will become manifest in a
$(2S+1)$-fold ground state degeneracy, corresponding to a high spin ($S$) state. For instance, if
in a system with $N=2$ electrons the lowest energy states in the subspaces $S_z=-1,0,1$ are the
degenerate ground states, then the $\frac12$ 
spins of two electrons point into the same direction, building an $S=1$ super spin. Because the
SU(2) symmetry of the individual electron spins is not broken, also this composite super spin has
full rotational symmetry.
The $S_z$ quantum numbers of the degenerate $S_z$ subspaces then correspond to the magnetization
of this composite spin system. Note that the spin-orbit interaction lowers the 
symmetry of the super spin, as it breaks the SU(2) invariance of the individual electron spins which
form the super spin.

For practical reasons, we extract the total spin quantum number $S$ of the ground state from its 
ground state degeneracy $(2S+1)$, which is obtained from the $S_z$ subspace ground state energies.
We have checked that this is equivalent to calculating the total spin $S$ of the ground state directly.

\subsection{Hubbard chain vs. edge states}

First we study the crossover from a usual Hubbard chain to interacting edge states. As explained 
above, the generalized model can be tuned continuously between these two limiting cases by means of 
the parameters $\lambda_{\rm us}$ and $\Gamma_1$. We postpone the analysis of backscattering to 
the next subsection and set $\lambda_{\rm bs}=1$ here.

\begin{figure}[!ht]
\centering
\includegraphics[width=\columnwidth]{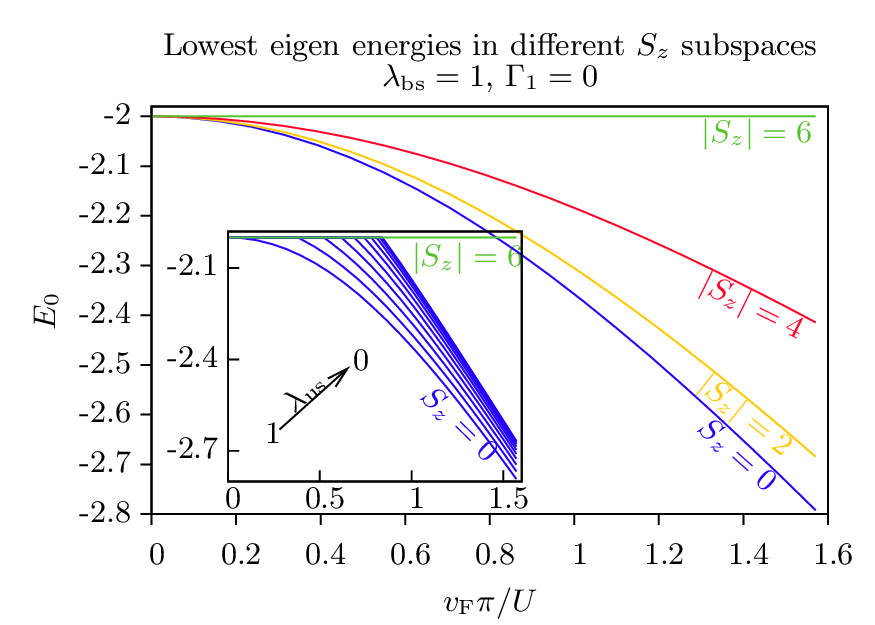}
\caption{(Color online) Lowest eigen energies in different $S_z$ subspaces for $N=12$ in the Hubbard 
limit with $\lambda_\text{bs}=1$, $\lambda_\text{us}=1$, $\Gamma_1=0$, and $U=1$. The inset shows 
the lowest eigen energies of the $S_z=0,6$ subspaces (zero and full spin-polarization) as the umklapp scattering is 
suppressed.  The lower lines correspond to $\lambda_{\rm us}=1$ and the higher 
lines to $\lambda_{\rm us}=0$. For the lines in between, $\lambda_{\rm us}$ decreases in steps of 0.2.}
\label{fig_hubbard_limit}
\end{figure}

It is most instructive to begin with the Hubbard chain limit of the generalized model, which is 
characterized by the full umklapp process strength $\lambda_{\rm us}=1$ and a suppressed 
momentum dependence of the interaction $\Gamma_1=0$. With this parameter set the direct model 
resembles a one-dimensional metal with a linear single particle dispersion instead of a cos-dispersion.
\footnote{We have checked that the linearization of the single particle spectrum does not change the 
results qualitatively. } The lowest eigen energies in the 
different $S_z$ subspaces for the parameter set described above are shown in Fig. \ref
{fig_hubbard_limit}. Obviously, the ground state is non-degenerate and resides in the $S_z=0$ 
subspace for arbitrary $v_F\pi/U$, except for the limit of infinitely large $U$. Thus, as expected,
no ferromagnetic phase transition exists for the Hubbard chain limit of the generalized model at
finite $v_F\pi/U$, in consistence with the Lieb-Mattis theorem\cite{lieb_mattis} which states that
the ground state of a system of one-dimensional interacting electrons 
has zero total spin and is non-degenerate with higher spin subspaces.
\footnote{The point $v_F\pi/U=0$ corresponds to a \emph{pathologic potential} in Ref.
\onlinecite{lieb_mattis}, as it can be reached by $U\rightarrow \infty$. At this point, all sectors
with total spin $0\leq S\leq S_\text{max}$ are degenerate.}

\begin{figure}[!ht]
\centering
\includegraphics[width=\columnwidth]{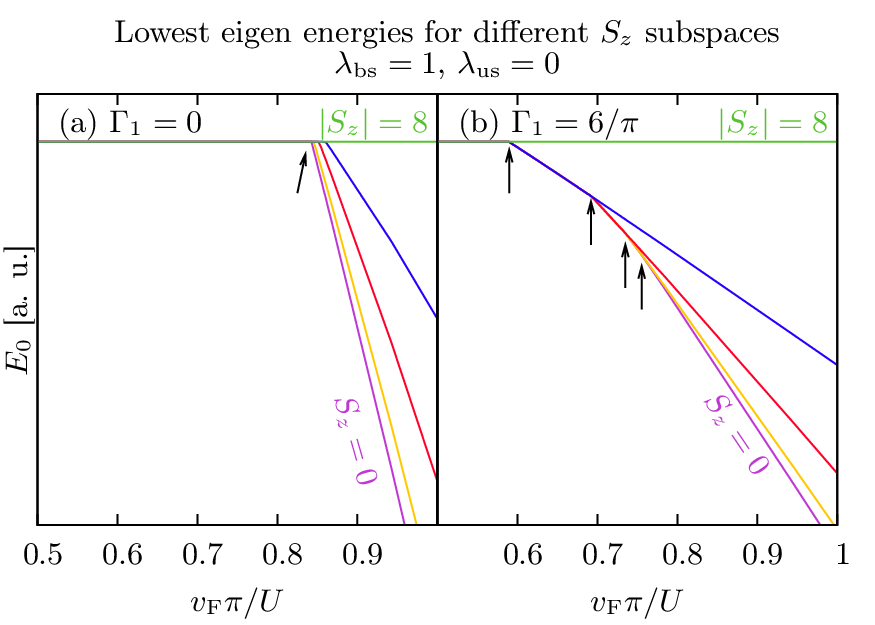}
\caption{(Color online) Lowest eigen energies with completely suppressed umklapp scattering $
\lambda_{\rm us}=0$ and different momentum-dependencies $\Gamma_1$. Furthermore, $N=16$, $
\lambda_{\rm bs}=1$, and $U=1$. Part (a) shows the case of a momentum-independent interaction 
vertex ($\Gamma_1=0$), where the ground state degeneracy jumps from 17 ($=2S_{\rm max}+1$) 
directly to 1 at the position indicated by the arrow. Part (b) shows the case of an interaction vertex with 
maximal momentum-dependence $\Gamma_1=6/\pi$. The arrows indicate a change in the ground state degeneracy.}
\label{fig_instant_jump}
\end{figure}

Next, the generalized model is tuned away from the Hubbard chain limit by suppressing the umklapp 
scattering $\lambda_{\rm us}<1$. Suppressed umklapp scattering is one of the properties of edge 
states which makes them fundamentally different from usual one-dimensional metals. In the inset of 
Fig. \ref{fig_hubbard_limit}, the lowest eigen energies of the $S_z=0$ and the $S_z=\pm N/2$ (full
spin-polarization) subspaces are shown as $\lambda_{\rm us}$ is reduced from 1 to 0 in steps of 0.2.
For 
any $\lambda_{\rm us}<1$ there is a nonzero critical value for $v_F\pi/U$ below which the lowest energy states 
of these two subspaces and also for all $S_z$ in between (not shown in the inset of Fig. \ref
{fig_hubbard_limit}) are equal. This corresponds to a high spin state of size $S=6$ in which the spins of 
all electrons point into the same direction. The critical point at which the transition between $S=0$ and 
$S=N/2$ takes place depends on the umklapp scattering strength 
\begin{equation}
\left[\frac{v_F\pi}{U}\right]^{(\Gamma_1=0)}_{\rm crit.} \propto \left(1 -  \lambda_{\rm us}\right)^y.
\end{equation}
For $N=12$ we find for the exponent $y\simeq0.5\pm0.02$. Obviously, the absence of umklapp 
scattering allows a high spin ground state. However, for $\lambda_{\rm us}<1$ and $\Gamma_1=0$, 
the system instantly jumps from zero polarization $S=0$ to the maximal possible polarization $S=N/2$ 
at the critical point $[v_F\pi/U]_{\rm crit.}^{(\Gamma_1=0)}$. This is a first order phase 
transition. For the case of completely suppressed umklapp scattering $\lambda_{\rm us}=0$ this is 
shown in Fig. \ref{fig_instant_jump} (a), where the lowest eigen energies of all subspaces are plotted as 
a function of $v_F\pi/U$: the $S_z=0$ subspace contains the non-degenerate ground state until at the 
critical point the lowest energy eigenstates of {\it all} subspaces form the degenerate ground state; no 
intermediate regime of $v_F\pi/U$ exists in which there is only a degeneracy between some of the 
$S_z$ subspaces.

The reason for the instant jump in the total spin is as follows: once the Stoner criterion $v_F\pi/U > [v_F
\pi/U]^{(\Gamma_1=0)}_{\rm crit.}$ is met, the interaction energy gain $\delta E_{U}(S)$ associated 
with developing a certain spin polarization $S$ is larger than the corresponding kinetic energy penalty $
\delta E_{\rm kin}(S)$. Unlike in two or three dimensions, however, for one-dimensional systems with 
momentum-independent interactions, $\delta E(S) = \delta E_U(S) + \delta E_{\rm kin}(S)$ has no 
minimum, i.e. $\delta E(S+1) < \delta E(S)$, for all $S<S_{\rm max}$. Thus, the system instantly 'flows' 
to the highest possible polarization $S_{\rm max}$, once the Stoner criterion is met. This is a rather 
common feature of one-dimensional systems with a constant interaction vertex (such as the Hubbard 
interaction) and can easily be observed in a variational calculation of the ground state properties (see 
Appendix \ref{appendix_mean_field}).

The momentum-dependence of the interaction vertex ($\Gamma_1>0$) reduces the interaction energy 
gain as the spin-polarization $S$ becomes larger. This is because for larger $S$, the Fermi level of the 
spin-up right-movers is shifted to higher momenta where the interaction is suppressed by the $S^{r=R}
_{k}$ factors [see Eq. (\ref{def_h_fs})]. Similarly, for the spin-up left-movers, the Fermi level is then 
shifted to smaller momenta, where the $S^{r=L}_k$ suppress the interaction.\footnote{Note that the 
increase in the interaction vertex for the spin-down electrons with lowered Fermi level is 
overcompensated by the suppression due to the higher Fermi level of the spin-up electrons, so that in 
total the interaction is reduced as $S$ grows.} As a result, $\delta E(S)$ develops a minimum at $S=S_
{\rm min}<S_{\rm max}$, and the system is stable there. Intuitively, this may be understood on the 
basis of a variational calculation (see Appendix \ref{appendix_mean_field}). Within exact 
diagonalization one finds that with $\Gamma_1=6/\pi$, the ground state degeneracy increases 
successively from 0 to $2S_{\rm max}+1$ by first adding the lowest energy eigenstates of the $S_z=
\pm1$ subspaces to the ground space, and then the $S_z=\pm2$ subspaces and so forth. This is 
shown in Fig. \ref{fig_instant_jump} (b).

\begin{figure}[!ht]
\centering
\includegraphics[width=\columnwidth]{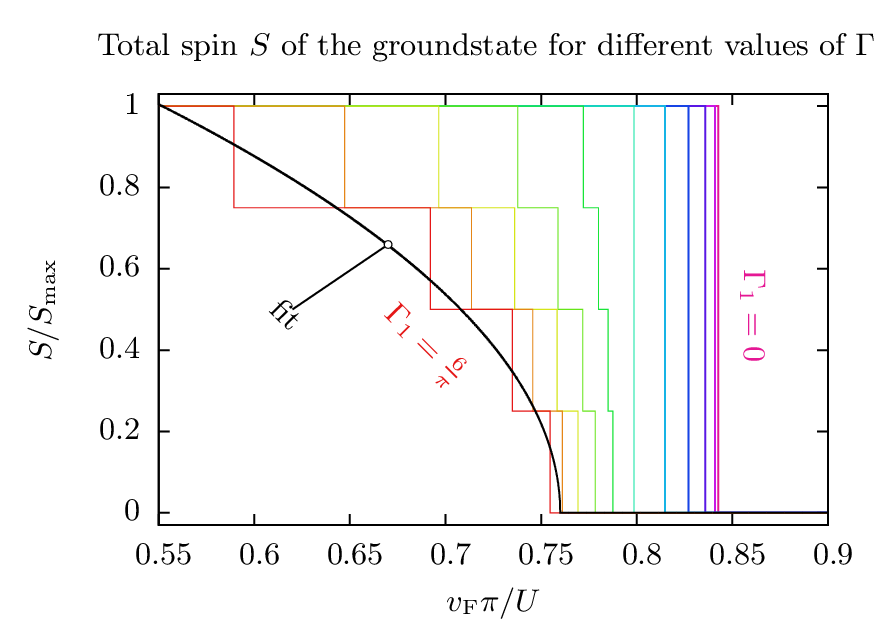}
\caption{(Color online) Dependence of the spin-polarization $S$ on $v_F\pi/U$ for different strengths of 
the momentum-dependence $\Gamma_1$ from $\Gamma_1=0$ (rightmost curve) to $\Gamma_1=
\frac{6}{\pi}$ (leftmost curve) in steps of $\Delta \Gamma_1 = \frac{6}{10\pi}$. The length of the edge is 
$L=48$, $\lambda_\text{bs}=1$ and $\lambda_\text{us}=0$. The smooth curve is a power law fit to the plateau centers of the $\Gamma_1=\frac6\pi$ steps with exponent $\beta=0.5$ (see text).}
\label{fig_g1_scan_mag_steps}
\end{figure}

\begin{figure}[!ht]
\centering
\includegraphics[width=\columnwidth]{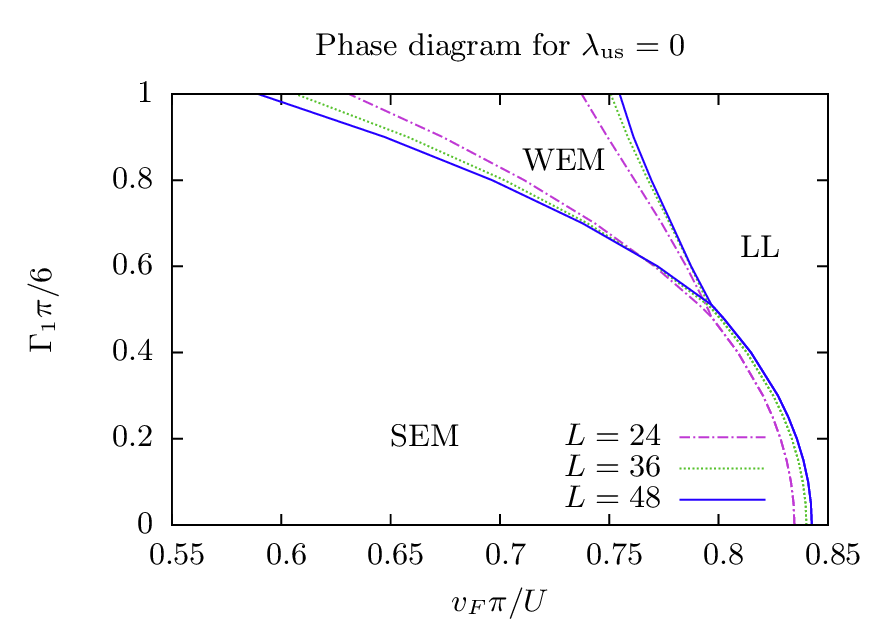}
\caption{(Color online) Phase diagram for lengths $L=24$, $L=36$ and $L=48$ . For small 
velocity dependence $\Gamma_1$ of the interaction, only the Luttinger liquid (LL) phase and the 
saturated edge magnetism (SEM) phase exist, whereas above the critical value of $\Gamma_1=
\Gamma_1^c$ the weak edge magnetism (WEM) phase appears. }
\label{fig_phase_diagram}
\end{figure}

If the total spin $S$ is plotted as a function of $v_F\pi/U$, $S$ decreases from $S_{\rm max}$ to 0 in 
steps. These steps correspond to the positions $v_F\pi/U$, where the degree of the 
ground state degeneracy changes, indicated by arrows in Fig. \ref{fig_instant_jump}. For $\Gamma_1=0$, there is one step where the spin-polarization 
jumps from $S=S_{\rm max}=N/2$ to $S=0$, while for the maximal $\Gamma_1=6/\pi$, there are $N/4$ 
steps at each of which the spin-polarization is decreased by $\Delta S=2$.\footnote{This decrease of 
$2$ in $S$ is due to the contribution of the left- and right-moving branch to the spin-polarization: each 
branch contributes one spin flip.} Fig. \ref{fig_g1_scan_mag_steps} shows these two limiting cases and 
how the steps evolve as $\Gamma_1$ is varied from 0 to $6/\pi$. The momentum-dependence must 
have a minimum strength $\Gamma_1>\Gamma_1^c\simeq 1$, in order to break the one big spin-
polarization step of height $N/2$ into many small steps of height 2. Thus, for $\Gamma_1 > 
\Gamma_1^c$ there is a regime of weak edge magnetism, meaning that the total spin $S$ of the ground state
is smaller than the maximal spin $S_\text{max}$, in addition to the usual saturated edge 
magnetism for small $v_F\pi/U$ (i.e. $S=S_\text{max}$) and the Luttinger liquid regime for large
$v_F\pi/U$ with $S=0$. Figure \ref {fig_phase_diagram} shows a diagram in which the phase boundaries
between the Luttinger liquid (LL), 
the saturated edge magnetism (SEM) and the novel weak edge magnetism (WEM) are shown for 
different system sizes $N=8,12,16$.

Note that the non-zero $\Gamma_1^c$ found in the exact diagonalization reveals a weakness of the 
fermionic mean-field theory in which this minimum momentum dependence, above which a WEM 
regime appears, is zero (see Appendix \ref{appendix_mean_field}). A non-zero $\Gamma_1^c$ means 
that the small momentum-dependencies which always follow from a dependence of the Bloch wave 
functions in usual one-dimensional conductors on the momentum are not necessarily sufficient to 
stabilize the weak edge magnetism; the momentum-dependence of the interaction vertex must be 
sufficiently strong for this.

In the limit $L\rightarrow\infty$, which cannot be accessed within exact diagonalization, of course, $S/
S_{\rm max}$ becomes a smooth function of $v_F\pi/U$. We approximate this smooth function by a 
power law
\begin{equation}
S/S_{\rm max} \sim \left[\left(\frac{v_F\pi}U\right)_{\rm crit.} - \frac{v_F\pi}U\right]^\beta.
\end{equation}
Fermionic mean-field theory (see Appendix \ref{appendix_mean_field}) predicts 
$\beta =0.5$. Because the exact diagonalization study is limited to small systems $N\leq 16$, it is difficult to obtain a decent estimate of the exact exponent $\beta$ within this work. Fitting the edge state limit of the generalized model to the center 
of the plateaus of the $N=16$ results of the exact diagonalization, we obtain $0.44<\beta<0.61$, dependent on how many plateaus are included in the fit. The critical point $\left(\frac{v_F\pi}U
\right)_{\rm crit.}\simeq 0.760$ for this fit is obtained by extrapolating the rightmost step from the data 
sets $N=4,8,12,16$ to $N=\infty$.

Interestingly, $\Gamma_1$ not only affects the order of the transition but also the critical $v_F\pi/U$. 
This is also not correctly predicted by the mean-field approach (see Appendix \ref
{appendix_mean_field}), which, independently of $\Gamma_1$, finds $v_F\pi/U=1$ to be the critical 
point. For small $\Gamma_1$ and $N=16$ the exact diagonalization gives
\begin{equation}
\left[\frac{v_F \pi}{U}\right]_{\rm crit.} \simeq 0.84 - 0.17 \Gamma_1^2.
\end{equation}

For the maximal $\Gamma_1=6/\pi$, the position of the leftmost step can be calculated by exact diagonalization for very large systems.\cite{large_system} We performed calculations for system sizes up to $L=180$ in order to extrapolate this step position. Within the limits of the accuracy of this extrapolation, the critical point $v_F\pi/U=0.5\pm 0.001$ between the SEM and the WEM regime coincides with the mean-field prediction (see Appendix \ref{appendix_mean_field}). This extrapolation to the thermodynamic limit, in combination with the extrapolation of the critical point between the WEM and the LL regime, is a strong evidence for the existence of the WEM phase for $0.5\leq v_F\pi/U \leq 0.760$ in the thermodynamic limit.

For completeness we note that our exact diagonalization analysis shows that a SEM phase also exists 
in the general model with umklapp scattering $\lambda_{\rm us}=1$ if $\Gamma_1>0$. However even 
for $\Gamma_1=6/\pi$ there is no weak edge magnetism phase between the Luttinger liquid and the 
saturated edge magnetism as long as $\lambda_{\rm us}=1$.

\subsection{The relevance of backscattering}

The backscattering Hamiltonian $H_1^{\rm bs}$ is important for the SU(2) invariance of the 
Hamiltonian. It is easily seen that only for $\lambda_{\rm bs}=1$ the SU(2) symmetry is preserved. At real graphene edges, of course, the backscattering cannot be tuned experimentally. Nevertheless it is 
interesting to study the consequences of a suppression of $H_1^{\rm bs}$ since in a bosonization 
treatment of the generalized model $H_1^{\rm bs}$ translates to a sine-Gordon term which is difficult to 
analyze. Therefore, some insight into the relevance of $H_1^{\rm bs}$ is helpful from a theoretical point of 
view. Following the philosophy of the previous subsection, we restrict the discussion to the spin 
polarization properties of the ground state. The analysis of more complicated observables such as spin-spin
correlation functions is beyond the scope of this work and will be discussed in another paper.

\begin{figure}[!ht]
\centering
\includegraphics[width=\columnwidth]{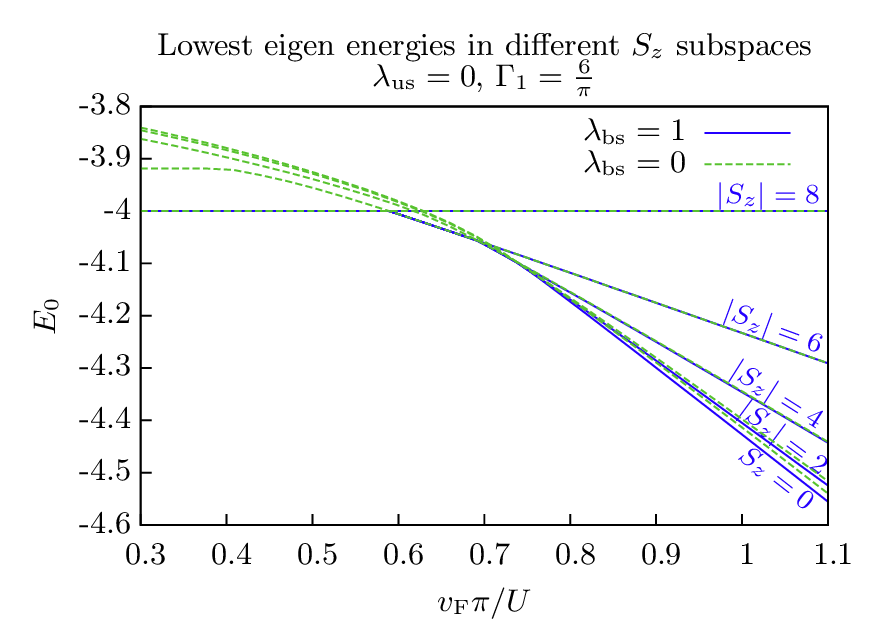}
\caption{(Color online) The lowest eigen energies of different $S_z$ subspaces with (solid blue)
 and without (dashed green) backscattering calculated for an edge of length $L=48$. The broken SU(2)
 symmetry in the case of $\lambda_\text{bs}\neq 1$ lifts the degeneracy of the ground states in the
 different $S_z$ subspaces. }
\label{fig_backscattering}
\end{figure}

Fig. \ref{fig_backscattering} compares the lowest eigen energies of the $S_z$ subspaces from 
calculations with and without backscattering. The most striking feature of the suppression of 
backscattering is the lifting of the ground state degeneracy in the SEM regime. This effect is easily 
understood by noting that the very reason for the ground state degeneracy in the $\lambda_{\rm bs}
=1$ case was the SU(2) symmetry, which, however, is broken for $\lambda_{\rm bs}<1$. Interestingly, 
the lifting of the degeneracy is such that the lowest energy states of the subspaces with highest $|S_z|
$, in the ground space for $\lambda_{\rm bs}=1$, form the ground state for $\lambda_{\rm bs}<1$. This 
means that suppressing backscattering introduces an Ising anisotropy along the spin quantization axis 
chosen in the definition of the model.

Apart from this degeneracy lifting, the evolution of the ground state properties with $v_F\pi/U$ is very 
similar for calculations with and without backscattering. The positions of the highest spin-polarization 
steps are practically unchanged. Only at the steps close to the phase transition between the LL and the 
WEM regime, a deviation of the $\lambda_{\rm bs}=0$ results from the $\lambda_{\rm bs}=1$ results 
can be observed. A handwaving explanation of this behavior can be given in terms of the bosonization 
analysis of the WEM regime in Ref. \onlinecite{tem_schmidt_loss_2010}. As soon as the Fermi levels 
for the up-spin electrons and the down-spin electrons are split, the backscattering process for electrons 
right at the Fermi surface is forbidden because it is not momentum-conserving. Thus, in order to 
conserve momentum, the electrons are forced to scatter to higher energies if there is a non-zero
spin-polarization. This mechanism suppresses the backscattering. In the bosonization language the 
backscattering Hamiltonian acquires a spatially oscillating phase which makes the corresponding 
operator irrelevant in the renormalization group. Thus in the WEM regime, not too close to the critical 
point, $H^{\rm bs}_1$ is suppressed and does not give an important contribution.

Close to the critical point, however, Fig. \ref{fig_backscattering} indicates that $H^{\rm bs}_1$ becomes 
more important. This observation is consistent with the qualitative bosonization argument: Close to the 
critical point the phase oscillations in the bosonic backscattering Hamiltonian get slower until they 
completely disappear at the critical point.

%% file: discussion.tex
On the basis of a generalized class of effective models for one-dimensional interacting electrons we
have studied the magnetic properties of a graphene zigzag edge. Using exact diagonalization we
confirmed the existence of three phases within these models, namely the saturated edge magnetism
phase
which is present at normal graphene edges, the Luttinger liquid phase which appears for edge states
with strongly enhanced bandwidth, and an intermediate regime of weak edge magnetism. The latter
phase is a realization of a ferromagnetic Luttinger liquid, a one-dimensional itinerant ferromagnet.
We presented evidence that the transition between the Luttinger liquid and the weak edge magnetism
phase becomes a second order quantum phase transition in the limit of long edges.

Beyond the identification of the magnetic properties of edge states, we examined the question why
electrons in one-dimensional edge states have such a rich phase diagram with two types of
ferromagnetic ground states, while usual one-dimensional electrons do not show any ferromagnetism.
In view of the Lieb-Mattis theorem,\cite{lieb_mattis} which actually forbids a spin-polarized ground state for
interacting electrons in one dimension, this question becomes even more pressing.

A closer inspection of the edge state model, which was derived directly from the graphene crystal
structure,\cite{tem_schmidt_loss_2010} revealed two unusual features of edge states which cannot be
found in other one-dimensional electronic systems, such as quantum wires, for instance. These are
(a) the total absence of umklapp processes in the effective electron-electron interaction,
independently of the filling factor, and (b) a strong momentum dependence of the effective
interaction vertex. Each of these features precludes the applicability of the Lieb-Mattis theorem. The momentum-dependence gives rise to a complicated non-local interaction, which cannot be written as $V(x,y)\hat n(x)\hat n(y)$, with $\hat n(x)$ an electron density operator, as it is required for the Lieb-Mattis theorem. And even in the limit of momentum-independent interactions ($\Gamma_1=0$ in the general model) the suppressed umklapp scattering makes the reformulation as a density-density interaction in real space impossible.
In order to further track down the particular consequences of these special features for the
magnetic properties of graphene edges, we replenished the direct model with a tunable umklapp
scattering term and replaced the interaction vertex by a generalized vertex function in which the
momentum-dependence can be switched on and off.

The study of this generalized model, which can be tuned continuously between its edge state limit
and a regime in which it describes normal one-dimensional metals, revealed the significance of the
two edge state features: the absence of umklapp processes is responsible for the existence of a
spin-polarized ground state, and the strong momentum dependence of the interaction vertex stabilizes
a regime of weak edge magnetism and gives rise to a second order phase transition between the
paramagnetic Luttinger liquid and the ferromagnetic Luttinger liquid.

It is interesting to note that the stabilization of the weak edge magnetism phase seems to be very
robust against changes in the details of the interaction vertex function. Apparently it is only
important that the vertex is suppressed as one of the four momenta of the participating fermions
gets close to one of the Dirac points. The exact functional form of this suppression, however, seems
to be irrelevant, since the qualitative behavior of the spin-polarization did not depend on whether
we used the interaction vertex of the direct model or the interaction vertex of the generalized
model with maximal momentum dependence. These two vertex functions have in common that they vanish
as one of the momenta approaches a Dirac point. However, their functional forms are very
different.

Finally we note that one-dimensional itinerant magnetism has also been studied in Hubbard chains with an additional second neighbor hopping,\cite{daul_1,daul_2} showing that it is indeed possible to define one-dimensional models which, at first sight, seem to comply with the prerequisites of the Lieb-Mattis theorem, but nevertheless have a high spin ground state. The physical picture behind the model discussed in Refs. \onlinecite{daul_1,daul_2}, however, is much different from the present work. Interestingly, the sign of the hopping amplitude to the nearest neighbor must be different from the sign of the next-nearest neighbor hopping for the system to have a ferromagnetic ground state.

Also, we emphasize that the model discussed here is the low-energy theory of a realistic system which may be studied experimentally. It has been derived in direct line from a two-dimensional lattice model of graphene/graphane interfaces.\cite{soi_schmidt_loss_2010,tem_schmidt_loss_2010}

%% file: mean_field.tex
We calculate the magnetic ground state properties of the generalized model within a fermionic
mean-field approximation. It is assumed that only the averages
$\left<c^\dagger_{kr\sigma}c_{kr\sigma}\right>$ are non-zero, so that the umklapp Hamiltonian
$H_1^{\rm us}$ and the backscattering Hamiltonian $H_1^{\rm bs}$ drop out of the mean-field
treatment. The resulting non-interacting Hamiltonian is diagonal in the momentum $k$, in the
direction of motion $r$ and in the $z$-spin projection, so that the mean-field theory is equivalent
to a variational ansatz based on the trial wave function
\begin{equation}
\left|M\right> = \prod_\sigma \left[\prod_{k<k_{F\sigma}} c^\dagger_{kR\sigma}\right]
\left[\prod_{k>-k_{F\sigma}} c^\dagger_{kL\sigma}\right] \left|0\right>
\end{equation}
with an asymmetric occupation of spin-up and spin-down states. The variational parameter $M\in[0,1]$
is related to the spin-dependent Fermi levels by
\begin{equation}
k_{F\sigma} = \sigma \frac\pi6 M,
\end{equation}
and to the spin-polarization $S$, used in Sect. \ref{ed_section}, by $M=S/S_{\rm max}$. For finite
size systems, as discussed in the main part of this paper, the Fermi level cannot be varied
continuously so that also $M$ is a discrete variable in this case. However, within mean-field theory
it is easy to perform the calculations in the thermodynamic limit, so that we will consider $M$ to
be a continuous variable and interpret it as the magnetization order parameter.

The variational energy $E(M)$ is easily calculated from the Hamiltonian $H$ in Eq.
(\ref{total_hamiltonian})
\begin{equation}
E(M) = \left<M\right| H \left|M\right> = \frac1{36} (\pi v_F -U) M^2 + \Gamma_1^2
\frac{U\pi^2}{5184} M^4.\label{variational_energy}
\end{equation}
For $U<\pi v_F$, the minimum of $E(M)$ is at $M=0$, while for $U>\pi v_F$ the mean-field ground
state has a finite magnetization
\begin{equation}
M= \min\left[\frac{\sqrt{72}}{\pi\Gamma_1}
\sqrt{1-\frac{v_F\pi}U},1\right]\label{mean_field_magnetization}
\end{equation}
Note that by definition the magnetization cannot become larger than 1. From Eq.
(\ref{mean_field_magnetization}) it becomes obvious that a non-zero momentum-dependence $\Gamma_1$
is required to stabilize the regime of weak edge magnetism. For $\Gamma_1=0$, the magnetization
would jump from 0 to 1 at the critical point $U=v_F\pi$. 

The existence of the weak edge magnetism can be traced back to the $M^4$ term in Eq.
(\ref{variational_energy}) which is generated by the momentum-dependence $\Gamma_1$ of the
interaction vertex. In dimensions higher than one, such $M^4$ terms emerge also from
momentum-independent interactions or directly from the kinetic energy, so that at least on the
mean-field level $\Gamma_1>0$ is required for the stabilization of weak ferromagnetism only in one
dimension.

%% file: direct_model_ed.tex
The direct model Hamiltonian $H^\text{dm}$ defined by Eqs. (\ref{eq:direct_model_H0} -
\ref{eq_effective_interaction_trig}) and the edge state limit of the generalized model Hamiltonian
$H$ [Eq. (\ref{total_hamiltonian}) with $\lambda_\text{bs}=1$, $\lambda_\text{us}=0$ and
$\Gamma_1=6/\pi$] are not exactly equal, as the general model linearizes the single particle
dispersion and replaces the factors $\mathcal{N}_p$ by the approximation $S_k^r$.
Nevertheless, the most important properties of graphene edge states, i.e. the momentum-dependence of
the interaction vertex and the absence of umklapp scattering, are properly described by both, the
direct model and the general model in the edge state limit.

\begin{figure}[!ht]
  \includegraphics[width=\columnwidth]{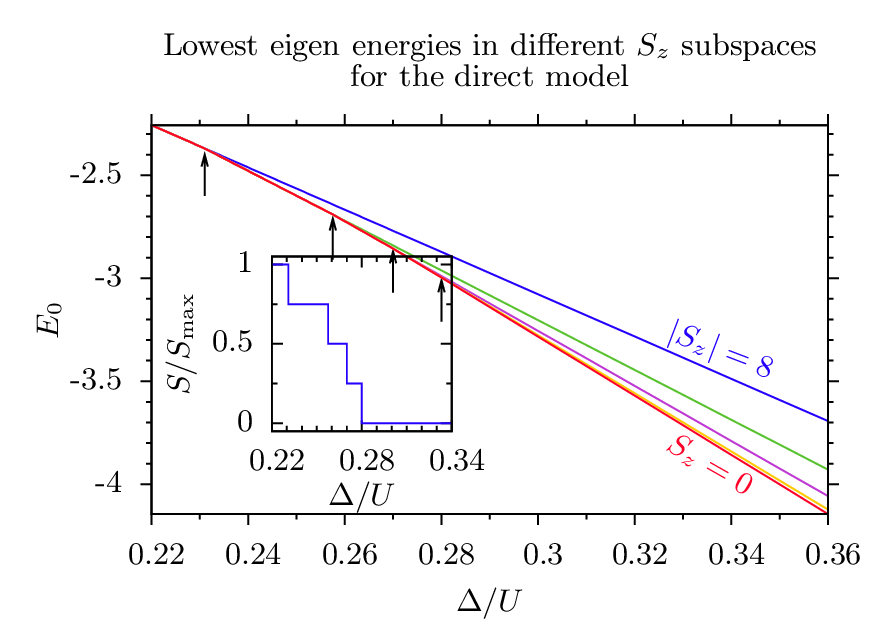}
  \caption{(Color online) Lowest eigen energies in the $S_z=0,\,\pm2,\, \pm4,\, \pm 6,\, \pm 8$
  subspaces (from bottom to top) for the direct model calculated for an edge of length $L=48$. 
  The inset shows the dependence of the spin-polarization $S$ as a function of $\Delta/U$ determined
  from the degeneracy of the ground state.
  \label{fig:direct_model}
  }
\end{figure}

In this appendix, we check that the direct model has qualitatively the same magnetic properties as
the general model in the edge state limit. In Fig. \ref{fig:direct_model}, we present the
spin-polarization $S$ as a function of $\Delta/U$, obtained from the exact diagonalization of the
direct model. The bandwidth parameter $\Delta$ of the direct model corresponds to the Fermi velocity
$v_F$ of the general model.

Clearly, for larger $\Delta/U$, which corresponds to the parameter $v_F \pi/U$ in the generalized
model, we obtain a Luttinger liquid phase with a ground state of total spin $S=0$. An
intermediate
regime with weak edge magnetism exists, where the total spin of the ground state $S<S_\text{max}$ is
not maximal. As in the exact diagonalization analysis of the general model in the main text, only some of the lowest
eigen energies in different
$S_z$ subspaces are degenerate and form the ground state.
For small $\Delta/U$, the saturated edge magnetism phase is reached and the spin of the
ground state is maximal, i.e., the lowest eigen energies in {\it all} $S_z$ subspaces are degenerate.

Note that the energy of the fully spin polarized eigenstate of the direct model Hamiltonian has a
finite slope (see Fig. \ref{fig:direct_model}). This is because the direct model lacks a symmetry of $H_0$ of the generalized model leading to
$E_0^{S_\text{max}}(v_F) = \text{const}$ (cf. Fig. \ref{fig_instant_jump}). As only the degeneracy of the lowest eigen energies are important, but not their absolute values, this difference does not have any physical consequences.

%% file: emw.bbl
\begin{thebibliography}{10}
\bibitem{large_system}%
  \BibitemOpen
  \bibinfo {note} {At $L/3$ filling, and $S_z=0$, we could access edges of
  length $L=48$ with a dimension of the corresponding Hilbert space of
  $2\,594\,212$ after exploiting all symmetries. For nearly maximal $S_z=L/6-2$
  and $L/3$ filling, we were able to calculate the groundstate energy of an
  edge with length $L=180$ with a Hilbert space dimension of only $26\,580$. In
  this case we were limited only by the 64 bit limit of the internal storage of
  the basis vectors and longer edges are in principle accessible by a
  modification of our code.}%
  \bibAnnoteFile{Stop}{large_system}%
\end{thebibliography}
